\documentclass[12pt,preprint]{aastex}
\usepackage{natbib}
\usepackage{amsmath,amsfonts,amssymb}
\usepackage{graphicx}
\def\idm#1{{\mbox{\scriptsize #1}}}

\newcommand\Chi{{(\chi^2_\nu)^{1/2}}}

\def\url#1{\texttt{#1}}

\shorttitle{Planet  Around HD 102272}
\shortauthors{Niedzielski et al.}
\begin{document}

\title{A Planet in a 0.6-AU Orbit Around the K0 Giant HD 102272}

\author{A. Niedzielski\altaffilmark{1,2}, K. Go\' zdziewski\altaffilmark{1},  A. Wolszczan\altaffilmark{2},  M. Konacki\altaffilmark{3,4},   G. Nowak\altaffilmark{1}, P. Zieli\'nski\altaffilmark{1} }

\altaffiltext{1}{Toru\'n Center for Astronomy, Nicolaus Copernicus University, ul. Gagarina  11, 87-100 Toru\'n, Poland, Andrzej.Niedzielski@astri.uni.torun.pl, Krzysztof.Gozdziewski@astri.uni.torun.pl, Grzegorz.Nowak@astri.uni.torun.pl, Pawel.Zielinski@astri.uni.torun.pl}
\altaffiltext{2}{Department of Astronomy and Astrophysics, Pennsylvania State University, 525 Davey Laboratory, University Park, PA 16802, alex@astro.psu.edu}
\altaffiltext{3}{Nicolaus Copernicus Astronomical Center, ul. Rabia\'nska 7,   87-100 Toru\'n, Poland, maciej@ncac.torun.pl}
\altaffiltext{4}{Astronomical Observatory, A. Mickiewicz University, ul. Sloneczna 36, 60-286 Pozna\'n,  Poland}

\begin{abstract}
We report the discovery of one or more planet-mass companions to the K0-giant HD 102272
with the Hobby-Eberly Telescope. In the absence of any correlation of the observed
periodicities with the standard
indicators of stellar activity, the observed radial velocity variations
are most plausibly explained in terms
of a Keplerian motion of at least one planet-mass body around the star. With the estimated
stellar mass of 1.9M$_\odot$, the minimum mass of the confirmed planet is 5.9M$_J$.
The planet's orbit
is characterized by a small but nonzero eccentricity of $e$=0.05 and
the semi-major axis of 0.61 AU, which makes it the most compact one discovered so far around GK-giants. This detection
adds to the existing evidence that, as predicted by theory, the minimum size of 
planetary orbits around intermediate-mass giants is affected by both planet formation processes and stellar evolution.
The currently available evidence for another planet around HD 102272 is insufficient to obtain an unambiguous two-orbit solution.

\end{abstract}


\section{Introduction}

Searches for planets around giant stars offer a unique way to extend studies of planetary
system formation and evolution to stellar masses substantially larger than 1 M$_{\odot}$ \citep{2003ApJ...597L.157S}.
These evolved stars have cool atmospheres and many
narrow spectral lines, which can be utilized in precision radial velocity (RV) measurements ($<$10 m s$^{-1}$). Searches for planets around giant
stars \citep[][and references therein]{2008arXiv0802.2590S, 2006A&A...457..335H, 2008A&A...480..215H, 2007ApJ...669.1354N} are beginning to provide the statistics, which are needed to constrain
the efficiency of planet formation as a function of stellar mass and chemical composition. In fact, the initial analyses
by \cite{2007A&A...472..657L} and \cite{2007ApJ...665..785J} suggest that the frequency of occurence
of massive planets is correlated with stellar mass. Because more massive
stars probably have more massive disks, these results appear to support
the core accretion scenario of planet formation \citep{2008ApJ...673..502K}. Furthermore, \cite{2008arXiv0801.3336P} 
have used the apparent lack of correlation between the frequency of planets
around giants and stellar metallicity to argue that this effect may imply
a pollution origin of the observed planet frequency - metallicity correlation
for main sequence stars \citep{2005ApJ...622.1102F}.

In time, the ongoing surveys  will also create an experimental basis with which to study 
the dynamics of planetary systems orbiting evolving stars \citep[e.g.][]{1996Sci...274..954R, 1998Icar..134..303D}.
Sufficiently large surveys of post-MS giants should furnish enough
planet detections to meaningfully address the problem of a long-term survival of
planetary systems around stars that are off the main-sequence (MS) and on their way to the white dwarf stage. In fact, the existing
data suggest a deficiency or absence of orbits with radii below 0.6-0.7 AU, which still awaits a fully satisfying
explanation \citep{2007ApJ...665..785J, 2008arXiv0802.2590S}. In addition, 
the recent detection of a planet around a post-red-giant phase
subdwarf V391 Peg \citep{2007Natur.449..189S} has already demonstrated that planets can survive both the RGB
and AGB phases of giant evolution.

In this paper, we describe the detection of a planet  around the K0-giant HD 102272 and discuss a possibility 
that at least one more planet-mass body exists in this system.
An outline of the observing procedure and description of the basic properties of HD 102272 are given in Section 2, 
followed by discussion of rotation and stellar activity indicators of the star in Section 3. 
The analysis of radial velocity 
measurements, is given in Section 4. Finally, our results are summarized  and further discussed in Section 5.

\section{Observations and properties of the star}

Our survey, the observing procedure, and data analysis have been described in detail elsewhere \citep{2007ApJ...669.1354N, 2008IAUS..249...43N}. 
Briefly, observations were made between 2004, July and 2008, January, with the Hobby-Eberly Telescope (HET) \citep{lwr98} equipped 
with the High Resolution Spectrograph (HRS) \citep{tull98} in the queue scheduled mode \citep{HetQ}. The spectrograph was used in the R=60,000 resolution  mode with a gas cell ($I_2$) inserted into the optical path, and it was fed with a 2 arcsec fiber. 

Radial velocities of HD 102272  were measured at 35 epochs spanning the period of about 1500 days.
Typically, the signal-to-noise ratio per resolution element in the spectra was 200-250 at 594 nm in 
9-16  minutes of integration, depending on the atmospheric conditions.  Radial velocities were measured 
using the standard I$_2$ cell calibration technique \cite{Butler+1996}. A template spectrum was 
constructed from a high-resolution Fourier Transform Spectrometer (FTS) I$_2$ spectrum and 
a high signal-to-noise stellar spectrum measured without the I$_2$ cell. Doppler shifts were 
derived from least-squares fits of template spectra to stellar spectra with the imprinted I$_2$ 
absorption lines. The radial velocity for each epoch was derived as a 
mean value of the independent measurements from the 17 usable echelle orders with
a typical uncertainty of  6-8 m s$^{-1}$  at 1$\sigma$-level. This RV precision made it quite sufficient to 
use the \cite{1980A&AS...41....1S} algorithm to refer the measured RVs to the Solar System barycenter.

HD 102272 (BD+14 2434) is a K0-giant 
with  V=8${^m}$.69, B-V=1${^m}$.02 and U-B=0${^m}$.69   \citep{1973A&AS...12..381} and  $\pi$= 2.76$\pm$1.11 mas 
\citep{hipparcos}.
 Its  atmospheric parameters were estimated using the method of \citet{2005PASJ...57...27T, 2005PASJ...57..109T}  as  T${_{eff}}$=4908$\pm$35 K, log(g)=3.07$\pm$0.12, and [Fe/H]=-0.26$\pm$0.08.
Comparing the star's position in the HR diagram with evolutionary tracks of \citet{girardi2000}, we estimate the mass and radius of HD 102272 to be M/M$_{\odot}$=1.9$\pm$0.3  and R/R${_\odot}$=10.1$\pm$4.6, respectively, from calibrations of \cite{alonso2000}.

\section{Analysis of the radial velocity data}

Radial velocity measurements of HD 102272 derived from the HRS spectra
 are shown in Figure 1. 
They reveal a correlated behavior of radial velocity variations of the star
on a timescale of about 130 days. Moreover, observations
made around MJD 54140 indicate that the amplitude of the RV curve may vary in time.

We have used both the nonlinear least-squares and the genetic (GA) algorithm
\citep{1995ApJS..101..309C} to model the observed RV variations with the standard, six-parameter
Keplerian orbits. A fixed 15 m s$^{-1}$ error was quadratically added to the 
formal RV uncertainities to account for any random RV variations intrinsic 
to the star, such as fluctuations of the stellar surface and possible solar-type oscillations. The anticipated value of this additional jitter was conservatively adopted to lie between 20 m s$^{-1}$, which is typical of stable K-giants \citep{2006A&A...454..943H}, and 5 m s$^{-1}$ as measured for stable dwarfs \citep{2005PASP..117..657W}. For example the solar-type oscillations alone, when extrapolated for HD 102272 as described by \cite{1995A&A...293...87K}, would account for a  6 m s$^{-1}$ RV variation. 
The 15 m s$^{-1}$ jitter is equivalent to f=0.005 spot on the surface of HD 102272  that would produce 2 m s$^{-1}$ BVS at the same time \citep{Hatzes2002}.

 The best-fit of a single orbit and its residuals are shown
in Figure 1. The fit, if interpreted in terms of an orbiting body, calls for
a $\sim$6 M$_{Jup}$ companion in a 127.6-day, slightly eccentric orbit around
the star (Table 1). The 0.61 A.U. radius of the orbit is even smaller than
the tightest planetary orbit around a red giant identified so far  \citep{2008arXiv0802.2590S}. However, not
 unexpectedly, the single planet model leaves highly correlated post-fit
residuals, indicating a possible presence of additional periodicities.

We have performed a search for the best-fitting two-companion Keplerian model
of the observed RV variations and identified several solutions of comparable
quality with very similar $\chi^2$ values of the fits. For example, the orbits
with the respective approximate periods of 179, 350, and 520 days and
eccentricities of 0.3, 0.5 and 0.7 are all almost indistinguishable in terms
of their goodness of fit. At the same time, the parameters of the 127.6-day
day orbit remain practically unchanged for all of the best two-orbit solutions.
An example of such a formal solution with the second orbit of 520 days is 
shown in Figure 1.

Because of a sparse sampling of the observed RV curve and the resulting
ambiguities in its modeling, it is useful to perform an additional test of
non-randomness of the residuals obtained from the best fit of a single
planet model to data. To accomplish this, we have chosen the test of
scrambled velocities \citep{Butler+2004} using the GA algorithm. In this test,
the residuals were randomly scrambled at the epochs of observations and
then searched for the best-fit parameters of the second orbit. A likelihood,
$p_{\idm{H}}$, 
that the residuals represent a white noise can be quantified as a ratio
of the number of best fits, for which $\chi^2$ is less than or equal to
the one derived from the fit to unscrambled data, to the total number of
trials. The result of the test with 100,000 trials is shown in Figure 2.
Clearly, the fit to the real signal stands apart from those for the randomly
scrambled data with $p_{\idm{H}}\sim 10^{-5}$, which strongly suggests
the presence of a second, non-random signal in the RV data.

In principle, the existing ambiguities in the modeling of the RV variations
in HD 102272 can be constrained by imposing the obvious requirement that 
the true two-planet solution is dynamically stable. A convenient way
to apply this constraint is to use the GAMP algorithm \citep{2006ApJ...645..688G}, in which
an N-body model of the observed RV curve has the $\chi^2$ criterion modified
by a penalty term for unstable orbits, and the MEGNO test \citep{2000A&AS..147..205C} is
applied to check the stability of a given two-planet solution. 

We have performed a GAMP search for the dynamically acceptable orbits of 
the second companion, with the MEGNO integration time set to $\sim$600
orbital periods of the outer companion. The results, in which the $\chi^2$
values relative to the best solution found by the search are mapped onto
the semi-major axis - eccentricity plane, are shown in Figure 3.

The plot reveals three regions of stability, one of which is located
between the mean motion resonances (MMR) of 5:2 and 3:1, another one is extended
along the 4:1 MMR, and yet another one occupies a high-eccentricity region
between the 4:1 and 5:1 MMRs. Further discussion of these results will be
given in the Chapter 5.

\section{Stellar photometry, rotation, and line bisector analysis}

We have followed the standard procedure to verify that the observed RV variations are due to orbital motion. It included an examination of the existing photometry of the star for possible variations and an analysis of the bisectors and curvatures of the selected spectral line profiles.  We have also searched for stellar acivity signatures in the H$\alpha$ line.

The most precise existing photometry of HD 102272 consists of 70 photometric observations of the star in the Hipparcos \citep{hipparcos} H$_{P}$ filter  between JD 2447877.95490  and 2448961.44322. The observed scatter in H$_{P}$ amounted to only 0.015 mag,   and no variability above this threshold was detectable. 
The star was also observed by the  Northern Sky Variability Survey (NSVS) \citep{2004AJ....127.2436W}. We have analyzed the 31 photometric observations made between JD=2451318.188426  and JD=2451630.300686, which were of a sufficient quality, and found no photometric variations of the star. 
Finally, our Lomb-Scargle periodogram analysis of these data did not reveal
any spectral peaks above a threshold level set at 6 times
the mean power computed for periods shorter than one month. Of course, it should be kept in mind that the Hipparcos and NSVS data were not simultaneous with our RV measurements.

The projected rotational velocity of HD 102272, $v sin i= $3 $\pm$ 1 km s$^{-1}$, was estimated using the cross-correlation method \citep{Benz+Mayor1984}. From this value and the adopted stellar radius we have obtained an estimate of the rotation period of P$_{rot}\sim$170 days, which is longer than the observed 127.6-day period of radial velocity variation and appears to be typical for K0 giants \citep{deMedeiros1996}. In principle, given the error estimates of \cite{alonso2000}, the rotation period of HD 102272 may range from 90 to 250 days.

Most analyses of the variations of spectral lines using line bisectors are based on a cross-correlation function (CCF) representation of an ``average'' spectral line of the observed star. As the $I_2$ lines affect stellar spectra in the iodine cell method of RV determination, they have to be properly removed before the CCF computation. An efficient method to accomplish this has been proposed by \citep{mf05}.

In our implementation of the method, we first cross-correlated the stellar spectrum with that of the flat-field $I_2$, in order to measure the wavelength offset between the iodine lines present in both spectra. The flat field flux was adjusted to the new wavelength scale, adding the previously determined offset by using a Hermite spline interpolation \citep{hill82}. 

The CCFs were computed by correlating the stellar spectrum with a numerical "zero-one", in which the non-zero points were aligned with the positions of stellar absorption lines at zero velocity. The numerical mask was constructed out of 980 lines of a very high SNR spectrum of HD 17092, the first star with a planet in our survey  \citep{2007ApJ...669.1354N}, cleaned from spectral features lying within $\pm \, 30 \; km \, s^{-1}$ of the known telluric lines \cite{Han03}. Out of the 35 resulting bisector measurements, three were affected by a low SNR in the original spectra and hence removed from further analysis. The typical uncertainties of the $BVS$, calculated with the expression given by \cite{mf05} were around $\sim$ 20 m s$^{-1}$. The mean value of BVS was calculated to be 25 $\pm$ 21 m s$^{-1}$. 

As both the Ca II K emission line (393.4 nm) and the infrared Ca II triplet lines at 849.8-854.2 nm are outside the range of our spectra, we used the H$\alpha$ line (656.28 nm) as a chromospheric activity indicator. To minimize a contamination from the telluric lines, we measured the EW of the central part of the line profile defined by I/I$_{c}$$\leq$0.7. Our analysis has shown
that the EW exhibited a 0.69\% rms scatter around its
mean value of $944.4\pm 6.5 m\AA$, which was not correlated with the observed RV variability (r = 0.06). 

The analysis of the existing photometry and bisector variability span for HD 102272 is summarized in Figure 4 which shows the Lomb-Scargle periodograms computed for all the time series of interest. We conclude that the only measured parameter, which exhibits a significant periodic variability is radial velocity. Similarly, as evident in the plot of BVS against the RVs in Figure 5, no correlation between the bisector velocity span and the radial velocities exists in HD 102272.

Using the scatter seen in the Hipparcos photometry of the star, and its rotational velocity determined above, we can estimate the amplitude of radial velocity variations and of the bisector velocity span due to a possible presence of a spot on the stellar surface \citep{Hatzes2002}. The observed  radial velocity amplitude of HD 102272 is almost an order of magnitude larger than 35 m s$^{-1}$ RV amplitude predicted by \cite{Hatzes2002} for a spot with  a filling factor of f=0.015, on a star rotating at $v sin i=$3 km s$^{-1}$. The expected  bisector variations of  5 m s$^{-1}$ are comparable to the precision of our radial velocities measurements and cannot have a detectable effect on our results. Similar results have been derived with the NSVS photometry using the spot-induced RV variation modeling described by \citep{sd97}.

In principle, it is  possible  that HD 102272, having no spots over the periods covered by both the Hipparcos data  ($\sim$1083 days) and those from the NSVS ($\sim$312 days), had developed a spot later on, at the time of our RV measurements. Using the results of  \cite{Hatzes2002} and \cite{sd97}, we estimate that, to generate RV variations as large as K=155 m s$^{-1}$, at a period similar to that of HD 102272 rotation, would require a large spot with the filling factor of f=0.17-0.22. Such a spot would translate into a 50-60  m s$^{-1}$ bisector velocity span that is most certainly excluded by our measurements. We also find it unlikely that such a large spot, apparently not present in the NSVS data mere $\sim$1400 days prior to our observations, would show no evolution in our measurements collected over 1500 days or 11 cycles of the observed periodicity. Therefore, we conclude that a rotating spot on the surface of HD 102272 is unlikely to be a cause of the observed periodic radial velocity variations. 

\section{Discussion}

Our observations of the K0-giant, HD 102272, reveal that the measured radial velocities of this star undergo
strictly periodic  variations at a period of 127.6$\pm$0.3  days. When interpreted in terms of a Keplerian motion, this periodicity indicates
the presence of a sub-stellar companion with a minimum mass of 5.9M$_J$, in a low
eccentricity, e=0.05$\pm$0.04 orbit, 0.61 A.U. away of the star (Table 1). In addition, our data indicate a possible presence of another, more
distant planetary-mass companion. Unfortunately, in this case, the sparse sampling of a putative second orbit does not allow an unambiguous distinction between orbital motion and intrinsic stellar effects. In any case, the inner planet has the shortest orbital period and the most compact orbit among planets around GK-giants discovered
so far \citep[][and references therein]{2008arXiv0802.2590S}.

As the long period RV variations in red giants may also be related to a combination of effects including
stellar rotation, activity, and non-radial pulsations \citep[e.g.][]{2006A&A...457..335H}, we have analyzed the photometric data
and the behavior of line bisectors of the star, following the established practices \citep{Quelozetal2001}. The details of our procedure
have been described by \citep{2008IAUS..249...49N}. Its application to HD 102272 has not shown any significant correlation between
the RV variations and the photometry or line variability of the star. Consequently, the most plausible explanation of our data is the presence of at least one sub-stellar mass companions around the star.

The orbital parameters of the confirmed planet are, within errors, independent of the choice of the orbit of a putative outer companion. The small orbital radius of the planet reinforces the existing evidence that orbits of GK-giant planets appear to be wider than
$\sim$0.6 A.U., as suggested by \citep{2007ApJ...665..785J, 2008arXiv0802.2590S}. These authors have considered the two obvious scenarios to create such a zone of avoidance around the GK-giants, namely the tidal capture of a nearby planet by the expanding star, and
a paucity of compact orbits around giants caused by peculiarities of the disk evolution around intermediate-mass stars
\citep{2007ApJ...660..845B}. In particular, simulations carried out by \citep{2008arXiv0802.2590S} indicate that, for giants which are over the peak
of the red giant branch, a tidal capture of planets at orbital radii $\leq$0.5 A.U. is possible. Clearly, much more observational evidence is needed to place these tentative conclusions on a firmer statistical footing.

Some of the past reports on planet discoveries around GK-giants do indicate a possible
existence of more planet-sized companions to the stars in question \citep{2008arXiv0802.2590S} , but none of them include analyses designed to
constrain their possible orbital parameters. As discussed above, and illustrated in Figure 3, our analysis has revealed three regions of plausible two-planet solutions. The better two of them, located between the MMRs of 5:2 and 3:1 and close to the 4:1 MMR (Table 1), offer a significant improvement of the quality of the fit, in comparison with the one-planet model. In addition,
these solutions remain dynamically stable over at least 1 Gyr, as shown by our numerical integration of the corresponding orbits. On the other hand, a more in-depth analysis of these orbits (Go\' zdziewski et al., in preparation) suggests that, because of the compactness of the stability regions for the two solutions, a probability for such systems to become trapped in the respective MMRs in the course of their dynamical evolution must be very low. Evidently, a confirmation or dismissal of a possible dynamical origin of the observed RV variations will have to wait until more data become available in the near future.

\acknowledgments
We thank  the HET resident astronomers and telescope operators for support. AN, AW and GN were 
supported in part by the Polish Ministry of Science and Higher Education grant 1P03D 007 30.
AW acknowledges a partial support from the NASA Astrobiology Program. 
 M.K. is supported by the Foundation for Polish Science through a FOCUS
grant.  K.G. was supported by the Polish Ministry of Sciences and Education grant 1P03D-021-29.
GN is a recipient of a graduate stipend of the Chairman of the Polish Academy of Sciences.
The Hobby-Eberly Telescope (HET) is a joint project of the University of Texas at Austin, the Pennsylvania State University, Stanford University, Ludwig-Maximilians-Universit\"at M\"unchen, and Georg-August-Universit\"at G\"ottingen. The HET is named in honor of its principal benefactors, William P. Hobby and Robert E. Eberly.

\clearpage

\begin{table}
\label{tab:table1}
\caption{
Orbital parameters of the two HD 102272 planets
derived from the best fit of a Keplerian model to the RV data.
${\cal M}$ is  the mean anomaly computed for 
the epoch of the first observation, $T_0=$~MJD~53042.976890.
}
\centering
\begin{tabular}{lll}
\hline
\hline
Parameter \hspace{1em}  
& HD~102272~{\bf b}  &  HD102272~{\bf c}   
\\
\hline
P~[days]                &   127.58 $\pm$  0.30   &   520 $\pm$  26 
\\
$T_0$~[MJD]             & 52146 $\pm$ 64   & 54135 $\pm$ 260
\\
K~[m\,s$^{-1}$]         &   155.5 $\pm$  5.6   &    59 $\pm$  11
\\
e                       &     0.05 $\pm$  0.04   &     0.68 $\pm$   0.06
\\
$\omega$ [deg]          &   118 $\pm$  58  &   320 $\pm$  10
\\
$m_2\sin~i$ [M$_{\idm{J}}$]     
                        &     5.9 $\pm$  0.2   &     2.6 $\pm$   0.4
\\
$a$ [AU] 		&     0.614 $\pm$  0.001   &     1.57 $\pm$   0.05  
\\
${\cal M}(T_0)$ [deg]
			&    12 $\pm$ 60   &   324 $\pm$  40 
\\
$V_0$ [m\,s$^{-1}$] 	& \multicolumn{2}{c}{-94.2 $\pm$ 3.6} 
\\
$\Chi$  		& \multicolumn{2}{c}{0.87}
\\
$\sigma_{RV}$~[m\,s$^{-1}$] 	& \multicolumn{2}{c}{15.4}
\\
\hline
\end{tabular}
\end{table}

\clearpage

%
%

\begin{figure}
\includegraphics[angle=0,scale=1.0]{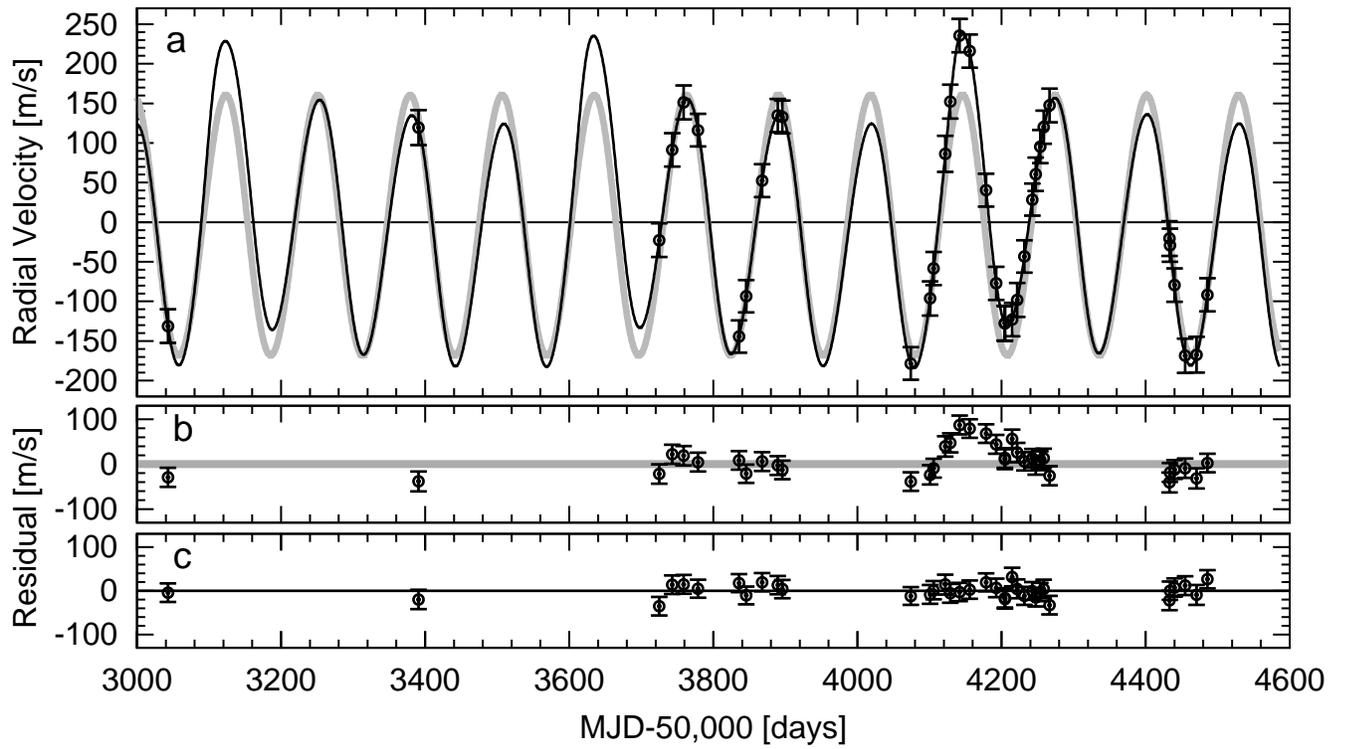}
\caption{(a) Radial velocity measurements of HD 102272 (filled circles) and
the best-fit models of a single orbit (gray line) and a two-planet system
close to the 4:1 MMR (solid line). (b-c) The respective best-fit residuals for
the above models.}
\end{figure}

\begin{figure}
\includegraphics[angle=0,scale=1.0]{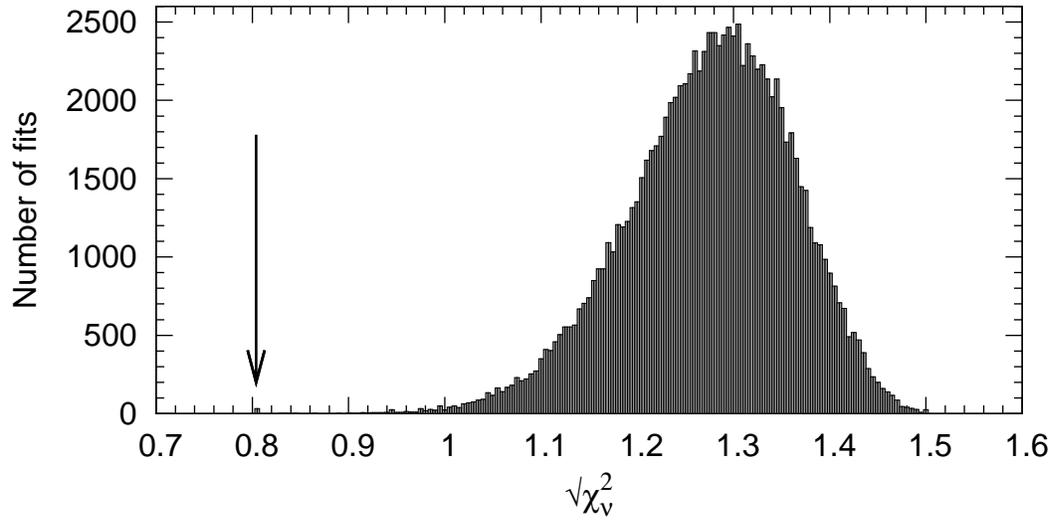}
\caption{A histogram of $\Chi$ for fits of a single-orbit, Keplerian model to 100,000 sets of RV residuals
scrambled with a randomly generated signal of the outer planet around HD 102272. 
Position of the best-fit solution for the real data is marked with an arrow.} 
\end{figure}

\begin{figure}
\includegraphics[angle=0,scale=1.0]{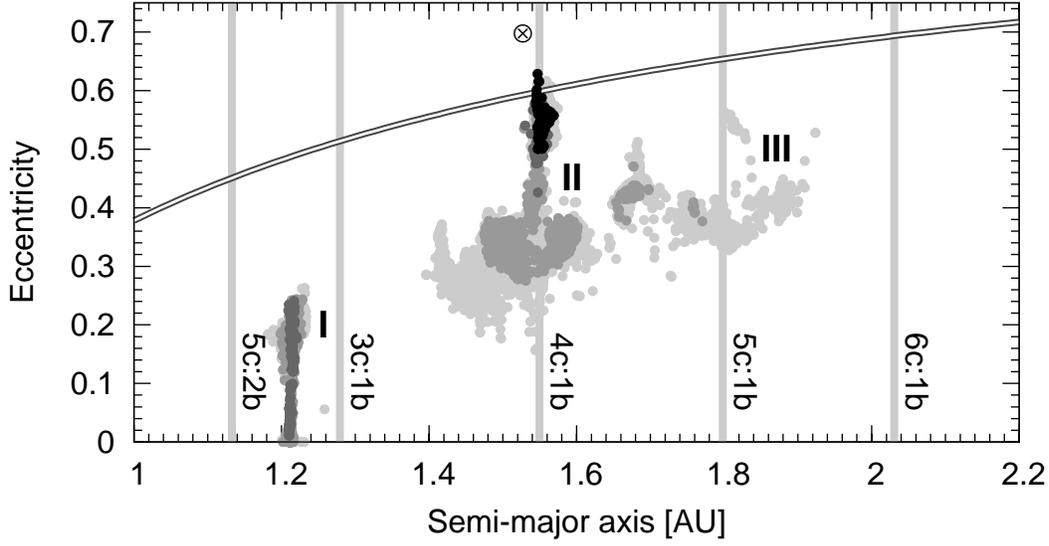}
\caption{An ensemble of Newtonian, best-fit models of a two-planet system 
generated with the GAMP code and projected onto the ($a_c,e_c$)-plane.
A location of each model in the diagram is marked with a grey-shaded circle.
The consecutive shades of grey (from the darkest to the lightest one) correspond to models with the values of $\Chi$ located
within 1$\sigma$, 2$\sigma$, and 3$\sigma$ of the $\Chi$ of the best-fit solution identified in the GAMP search, respectively.
Vertical lines denote positions of low-order mean motion resonances
between planets~b and~c. 
The double-lined curve marks the so called collision line of orbits defined by a condition: 
$a_b (1+e_b) = a_c (1-e_c)$. 
As a reference, the best-fit Newtonian model obtained without
stability constraints is marked with a crossed circle. See text for details.} 
\end{figure}

\begin{figure}
\includegraphics[angle=0,scale=1.0]{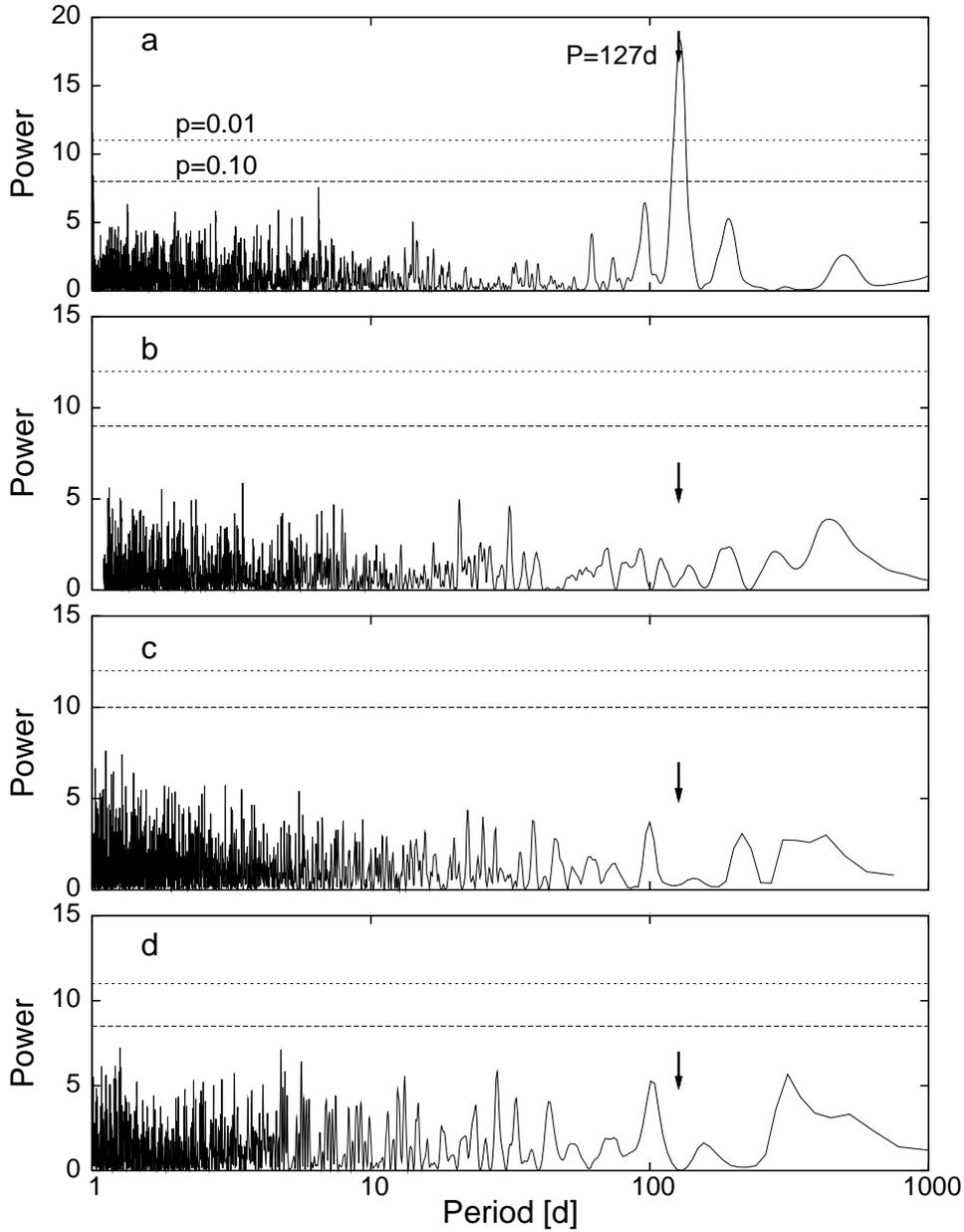}
\caption{Lomb-Scargle periodograms of (a) radial velocities, (b) bisector velocity span, (c) Hipparcos H$_p$ photometry, and (d) NSVS photometry. The position of the periodicity identified in the RV data is indicated with a vertical arrow. Two false alarm probability levels, 10$\%$ and 1$\%$ are marked with dotted lines.}
\end{figure}

\begin{figure}
\includegraphics[angle=0,scale=1.0]{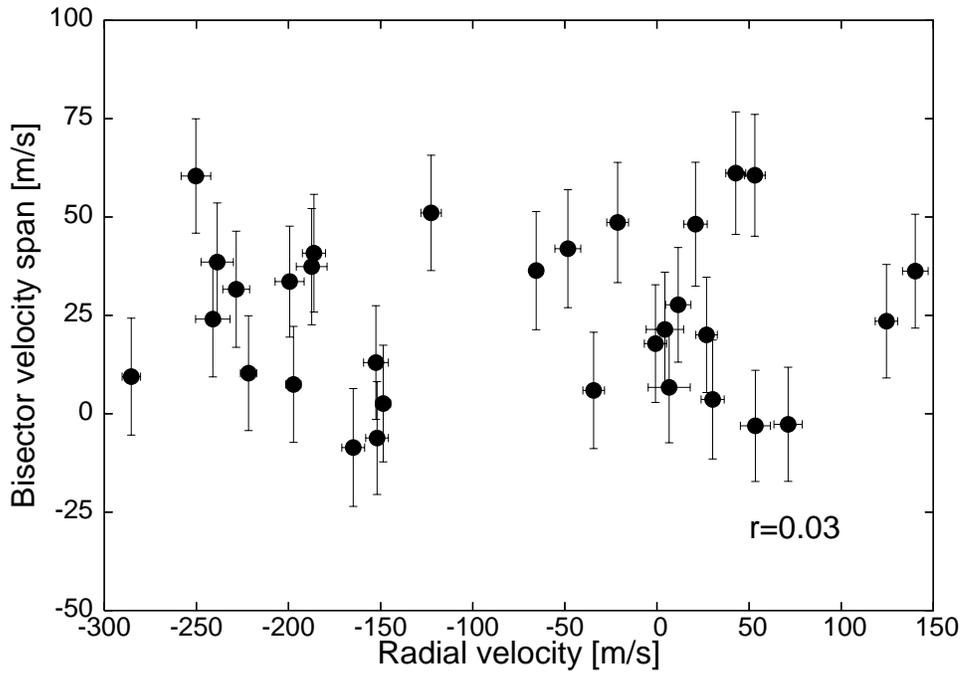}
\caption{ Radial velocity measurements of HD 102272 plotted against the bisector velocity span. Within errors, the BVS remains constant during the period covered by observations at the mean value of 25 m s$^{1}$ and rms scatter of 1$\sigma =$ 21 m s$^{-1}$. There is no correlation between RV and BVS (r=0.03).}
\end{figure}

\end{document}